\documentclass[11pt,a4paper,
]{article}

\usepackage{latexsym}
\usepackage{amsmath}
\usepackage{amssymb}
\usepackage{listings}
\usepackage{color}
\usepackage{graphicx}          
\usepackage{balance}

\usepackage{geometry}
\usepackage{acronym}

\newacro{IFS}{Interface File System}
\newacro{LIN}{Local Interconnect Network}
\newacro{XML}{eXtended Markup Language}
\newacro{DSP}{Digital Signal Processor}
\newacro{SPLIF}{Service Providing Linking InterFace}
\newacro{SRLIF}{Service Requesting Linking InterFace}
\newacro{OMG}{Object Management Group}
\newacro{TDMA}{Time Division Multiple Access}

\def\ie{i.\,e.,~}
\def\eg{e.\,g.,~}
\def\lqq{\lq\lq}
\def\rqq{\rq\rq}
\def\dq#1{\lqq #1\rqq}
\def\TTPA{TTP/A}

 \usepackage{eso-pic}
 \usepackage{color}
 \definecolor{DisclaimerGray}{gray}{0.92}

 \AddToShipoutPicture*{\put(40,800){\colorbox{DisclaimerGray}{\parbox{18.5cm}
 {\footnotesize \copyright IEEE, 2006. This is the author's version of the
 work. Personal use of this material is permitted. However, permission to
 reprint/republish this material for advertising or promotional purpose or for
 creating new collective works for resale or redistribution to servers or lists,
 or to reuse any copyrighted component of this work in other works must be
 obtained from the copyright holder. The definite version is published at
 IEEE Transactions on Industrial Informatics, 2(3):192–199, 2006.}}}}

\begin{document}

\lstset{
        extendedchars=true,
        basicstyle=\footnotesize\ttfamily,
         lineskip=2pt,
        }

\title{Time-Triggered Smart Transducer Networks}

\author{Wilfried Elmenreich\\
    wilfried.elmenreich@aau.at}

\date{IEEE Trans. on Industrial Informatics, 2(3):192–199, 2006}

\maketitle

\begin{abstract}
The time-triggered approach is a well-suited approach for building
distributed hard real-time systems. Since many applications of
transducer networks have real-time requirements, a time-triggered
communication interface for smart transducers is desirable,
however such a time-triggered interface must still support
features for monitoring, maintenance, plug-and-play, etc.

The approach of the \ac{OMG} Smart Transducer Interface consists of
clusters of time-triggered smart transducer nodes that contain
special interfaces supporting configuration, diagnostics, and
maintenance without affecting the deterministic real-time
communication. This paper discusses the applicability of the
time-triggered approach for smart transducer networks and presents a
case study application of a time-triggered smart transducer network.
\end{abstract}


\section{Introduction} \label{sec:introduction}

With the advent of modern microcontrollers it became feasible to
build low-cost smart transducers by equipping sensors and
actuators with a microcontroller and a standard network interface.
Several smart transducers are connected to a cluster using a
standard or non-standard fieldbus network. Many applications that
interact with the environment via transducers have real-time
requirements, that is to make correct actions at the right time.
So there is a need for an appropriate real-time communication
interface for smart transducers.

There are two major design paradigms for implementing distributed
real-time systems, the event-triggered and the time-triggered
approach. Simplified, an event triggered system follows the
principle of reaction on demand. In such systems, the environment
enforces temporal control onto the system in an unpredictable
manner (interrupts), with all the undesirable problems of jitter,
missing precise temporal specification of interfaces and
membership, scheduling etc. On the other hand, the event-triggered
approach is well-suited for sporadic action/data, low-power sleep
modes, and best-effort soft real-time systems with high
utilization of resources. Event-triggered systems do not ideally
cope with the demands for predictability, determinism, and
guaranteed latencies -- requirements that must be met in a hard
real-time system.
Time-triggered systems derive all triggers for communication,
computation, sensing and control by the global progression of
time. In this approach the concept of time is prevalent in the
problem statement as well as in the provided solution.

The objective of this paper is to present a time-triggered
approach for smart transducer networks that supports the hard
real-time requirements of embedded applications while still
providing features for maintenance and plug-and-play.

The remaining parts of this paper are structured as follows: The
next section identifies basic requirements for smart transducer
networks. Section~\ref{sec:time-triggered} depicts the generic model
of a time-triggered system.
Section~\ref{sec:omg-transducer-standard} describes the \ac{OMG}
Smart Transducer Standard, which incorporates a time-triggered
communication interface to smart transducers while fulfilling the
requirements for a smart transducer. Section~\ref{sec:relatedwork}
gives an overview on related work on time-triggered smart transducer
networks. Section~\ref{sec:conclusion} concludes the paper.


\section{Requirements for Smart Transducers}

We make the following assumption: The sensors and actuators in the
system are distributed and implemented as smart transducers with a
network interface. The network connects these smart transducers to
a communication system with broadcast characteristics (\eg bus or
star topology). The network also contains local intelligence, \eg
for feedback control purposes or sensor information processing.
This intelligence is either implemented in the processing unit of
the smart transducers or provided by separate control nodes.

Large transducer networks will be divided into clusters where each
cluster connects a set of transducers via a bus. A gateway node
exports the interfaces of the nodes in the cluster to a backbone
network.

We have identified the following requirements for smart
transducers in such a network:

\begin{description}

\item[Real-Time Operation:] Most applications for transducers,
especially in the fields of process automation, automotive and
avionic networks, require timely actions, \eg for information
gathering, sensor processing and actuator setting. Thus, a smart
transducer should provide a real-time interface that allows for
such a coordination.

\item[Complexity management:] The number of sensors and actuators
employed in a typical system has drastically increased in the last
two decades. Thus, a smart transducer should provide means to
manage the system complexity when composing or changing a network
of transducers, \eg by supporting electronic datasheets.
Electronic datasheets contain a machine-readable self-description
of the transducer which can be used to support a
plug-and-play-like computer aided
configuration~(cf.~\cite{eccles:98}).

\item[Maintenance support:] Systems which are in operation for an
extended period of time usually require maintenance access to
smart transducers, \eg for reading sensor logs, calibration or
trimming of the sensor's output.

Often, the information to be monitored is not fully covered by the
data exchanged via the real-time interface. Therefore, the
monitoring operation requires an extra data channel for
communication of these additional data. In this case, it is
required that the real-time traffic among the transducers is not
affected by the monitoring operation in order to avoid a \dq{Probe
Effect}~\cite{gait:86,mcdowell:89} on the system.

An appropriate interface is essential for supporting effective
maintenance methods such as Condition-Based
Maintenance~\cite{bengtsson:04} where affected components are
repaired or replaced before their failure causes greater costs.

\item[Deterministic Behavior:] A system is deterministic, if a
given set of inputs always leads to the same system output.
Determinism is especially important if replicated transducers are
used to enhance the dependability of an application. For real-time
systems, a deterministic system must, for a given set of inputs,
always produce the same output with regard to values and timing.

Since transducers operate on the borderline to the analog process
environment, determinism is difficult to achieve. For example, in
the case of a sensor, the input comes from the process
environment, which is an analog system. Thus, even a digital
sensor will not be exactly value deterministic due to
digitalization and intrinsic sensor errors. If a system contains
such sources of indeterminism, consistency must by achieved by
mechanisms like inexact voting~\cite{par:92}, sensor
fusion~\cite{elmenreich:sensorfusionintro}, and sparse
time~\cite{kopetz:2001-22}.

\end{description}

\section{The Time-Triggered Approach} \label{sec:time-triggered}

The core mechanism of a time-triggered system is very simple. A
global schedule defines for each node which action it has to take
at a given point in time.

\begin{table*}[t]
\centerline{
\begin{tabular}{|c|c|c|c|c|}
  \hline
  Time & [0:00,3:00) & [3:00,5:00) & [5:00,8:00) & [8:00, 12:00) \\
  \hline
  Node A & send & receive & receive & execute task \\
  Node B & receive & send & receive & send \\
  Node C & receive & receive & execute task & receive \\
  Node D & receive & receive & idle & receive \\
  \hline
\end{tabular}}
  \caption{Example for a time-triggered schedule}
  \label{table:ttschedule}
\end{table*}

\begin{figure*}[t]
 \centerline{\includegraphics[width=.8\textwidth]{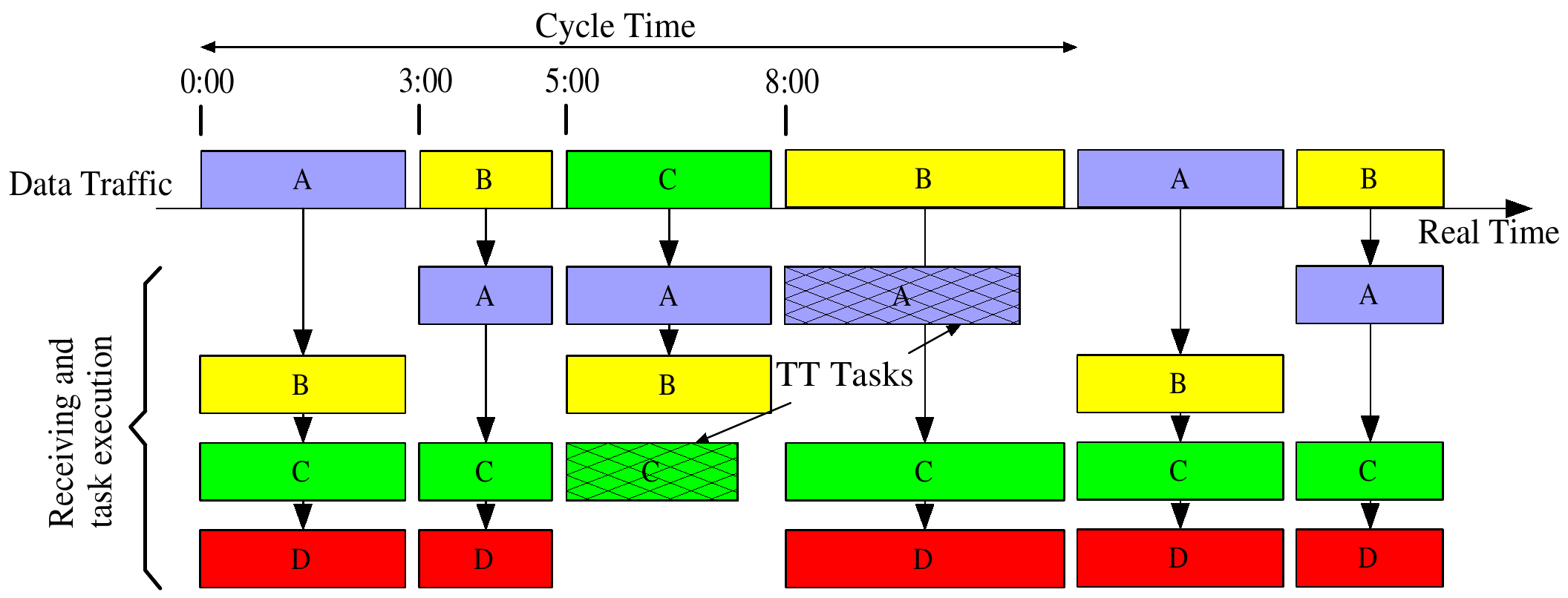}}
  \caption{Execution of time-triggered schedule}
  \label{fig:ttschedule}
\end{figure*}
This schedule is executed periodically. A prerequisite for
time-triggered systems is that all communication partners agree on
the current execution state of the schedule and that the duration
of all communication and computation activities are bounded and an
upper bound for this duration is known.

An example for such a schedule is given in
Table~\ref{table:ttschedule}. Within a cycle, depicted by the {\em
cycle time}, each message is scheduled at a predefined point in
time -- in this example, a message from node A is scheduled at
time 0:00, a message from node B at 3:00, a message from node C is
scheduled at 5:00 and another message from node B is scheduled at
8:00. Figure~\ref{fig:ttschedule} depicts the execution of this
schedule.


For a network of transducers, the time-triggered approach comes
with the following advantages:

\begin{description}

\item[Global time:] The global synchronized time is a requirement
and a feature in time-triggered systems. The global time must be
established by periodic clock synchronization in order to enable
time-triggered communication and
computation~\cite[p.\,52]{kopetz:97}.

In case of a smart transducer, the global time provides each
measurement with a timestamp that can be globally interpreted.

\item[Conflict-free bus arbitration:] Time-triggered systems need
no explicit arbitration mechanism for bus access, since all
communication actions are scheduled at predefined points in time.
This simplifies communication design by making it possible to omit
message retry mechanisms and special data encodings for detecting
data collisions on the bus. Moreover, the electrical specification
of the bus system is not required to cover the case of multiple
partners concurrently accessing the communication medium as
senders.\cite[p.\,176]{kopetz:97}

\item[Synchronization of distributed actions:] The time-triggered
schedule allows a precise coordination of actions in the network.
Examples for such synchronization actions are:

\begin{itemize}

\item Synchronous measurements by several distributed sensors: If
the measured variable is a fast moving real-time value,
unsynchronized measurements from multiple sensors will lead to
significantly different results.

\item Cascaded measurements to avoid interference: Sensors that
emit an active signal, for example ultrasonic distance sensors,
may interfere with each other if the measurement is started
concurrently.

\item Synchronous actuating: Applications where two or more
actuators are manipulating the same process might require
synchronous action. An example for this case is an application
with multiple servos applied to the same shaft, where an
unsynchronized execution leads to increased electrical current
flow and load for the servo that actuates first.

\end{itemize}

\item[Determinism:] Because of the time-triggered coordination,
sources of indeterminism like race conditions are removed by
design. Time-triggered systems are therefore deterministic in the
value and in the time domain. This factor is especially important
in the case of replicated systems where voting is used on the
outputs to enhance dependability (cf. replica
determinsm~\cite{poledna:diss}).

\end{description}

Note that time-triggered communication alone, \ie using a \ac{TDMA}
communication scheme, is not sufficient to establish a
time-triggered architecture. For example in the \ac{LIN}
system~\cite{linspec}, the master follows a time-triggered
communication schedule, while the interface to the \ac{LIN} nodes
operates on a polling principle. Thus, \ac{LIN} does not support
synchronized measurements of multiple sensors within one cluster.

The most prominent example for a time-triggered approach is the
Time-Triggered Architecture~\cite{kopetz:2001-22}, which provides
a highly dependable real-time communication service with a
fault-tolerant clock synchronization scheme and error detection of
faulty nodes. This architecture is suitable to build
ultra-dependable computer systems for safety-critical
applications, where a mean-time-to-failure (MTTF) of better than
$10^{9}$ hours is required~\cite{suri:95,kopetz:faulthypothesis}.

\section{OMG Smart Transducer Standard} \label{sec:omg-transducer-standard}

In December 2000 the \ac{OMG} called for a proposal of a Smart
Transducer Interface (STI)
standard~\cite{omg:SmartTransducerInterfaceRFP}. In response, a new
standard has been proposed that comprises a time-triggered transport
service within the distributed smart transducer network and a
well-defined interface to a CORBA (Common Object Request Broker
Architecture) environment. The key feature of the STI is the concept
of an \ac{IFS} that contains all relevant transducer data. This
\ac{IFS} allows different views of a system, namely a real-time
service view, a diagnostic and management view, and a configuration
and planning view. The interface concept encompasses a communication
model for transparent time-triggered communication. This STI
standard has been finalized by the \ac{OMG} in January
2003~\cite{omg:SmartTransducerInterface1.0}.

\begin{figure}[htb]
\begin{center}
  \includegraphics[width=.5\columnwidth]{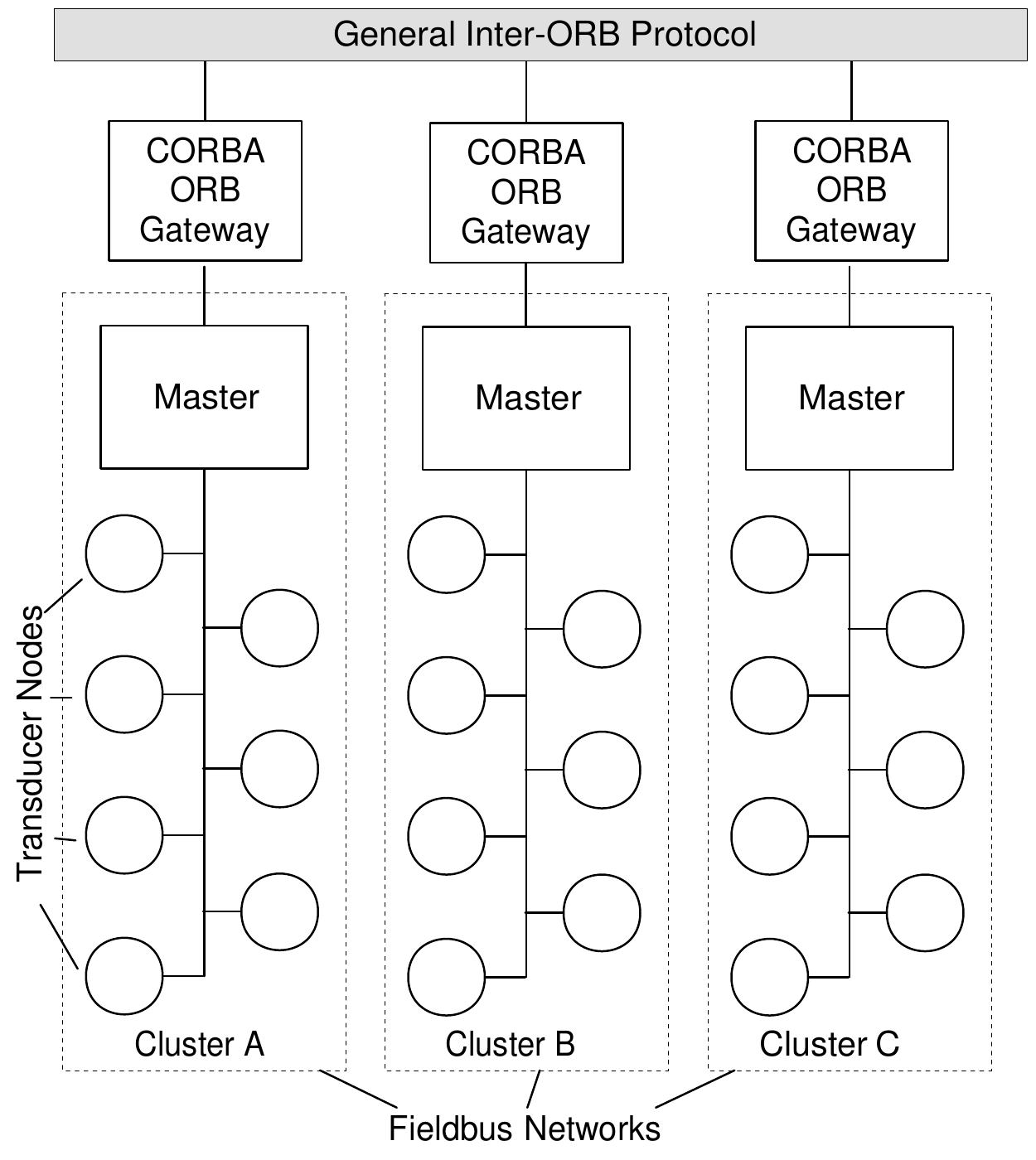}
  \caption{Multi-Cluster Architecture with CORBA Gateway}
  \label{fig:starchitecture}
\end{center}
\end{figure}

The STI standard defines a smart transducer system as a system
comprising of several clusters with transducer nodes connected to
a bus. Via a master node, each cluster is connected to a CORBA
gateway. The master nodes of each cluster share a synchronized
time that supports coordinated actions (\eg synchronized
measurements) over transducer nodes in several clusters.
Each cluster can address up to 250 smart transducers that
communicate via a cluster-wide broadcast communication channel.
There may be redundant shadow masters to support fault tolerance.
One active master controls the communication within a cluster (in
the following sections the term master refers to the active master
unless stated otherwise). Since smart transducers are controlled
by the master, they are called slave nodes.
Figure~\ref{fig:starchitecture} depicts an example for such a
smart transducer system.

It is possible to monitor the smart transducer system via the
CORBA interface without disturbing the real-time traffic.

The STI standard is very flexible concerning the hardware
requirements for smart transducer nodes, since it only requires a
minimum agreed set of services for a smart transducer
implementation, thus supporting low-cost implementations of smart
transducers, while allowing optional implementation of additional
standard features.

The information transfer between a smart transducer and its client
is achieved by sharing information that is contained in an internal
\ac{IFS}, which is encapsulated in each smart transducer.

\paragraph{Interface File System.}
The \ac{IFS}~\cite{holzmann:01} provides a unique addressing scheme
to all relevant data in the smart transducer network, \ie transducer
data, configuration data, self-describing information, and internal
state reports of a smart transducer. The values that are mapped into
the \ac{IFS} are organized in a static file structure that is
organized hierarchically representing the network structure
(Table~\ref{table:ifsaddress}).

\begin{table*}
\centerline{
\begin{tabular}{|p{.28\textwidth}|p{.15\textwidth}|p{.55\textwidth}|}
  \hline
  Element & Size & Description \\ \hline
  Cluster name  & 8 bit & Identifies a particular cluster. Native communication
(without routing) among nodes is only possible within the same
cluster. \\
  Node alias    & 8 bit & The node alias or logical name selects a
particular node. Some values have an associated special function,
\eg alias 0 addresses all nodes of a cluster in a broadcast
manner\\
  File name     & 6 bit & The file name addresses a certain file within a
node. A subset of files, the system files, have a special meaning in
all nodes. Each service of a node is mapped onto a file containing
sections for the service providing and service requesting linking
interface as well as for configuration/planning and
diagnosis/management data.\\
  Record number & 8 bit & Each file has a statically assigned number
of records. The record number addresses the record within the
selected file. Each record contains 4 data bytes. Note that each
file contains only the necessary number of records, thus, the
number of addressable records is statically defined for each file.\\
\hline
\end{tabular}}
  \caption{Hierarchical structure of an \ac{IFS} address}
  \label{table:ifsaddress}
\end{table*}

\paragraph{Communication via temporal firewalls.}

\begin{figure}[htb]
  \centerline{\includegraphics[width=.7\columnwidth]{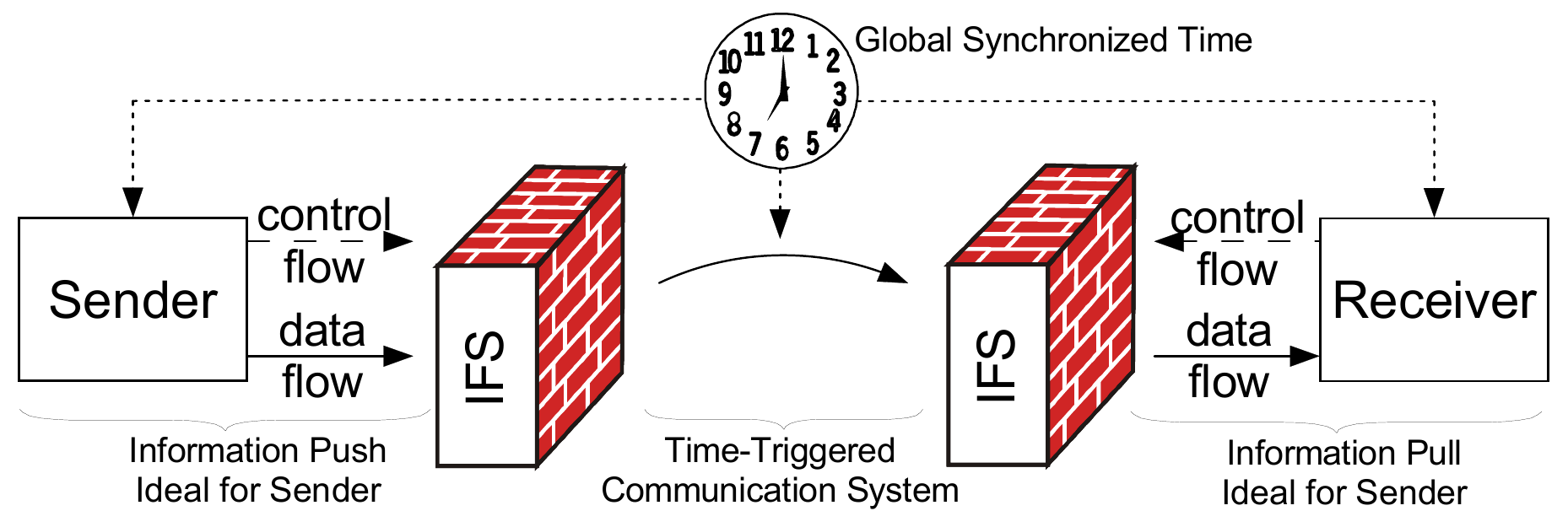}}
  \caption{Temporal Firewall}
  \label{fig:temporalfirewall}
\end{figure}

A time-triggered sensor bus will perform a periodical time-triggered
communication to copy data from the \ac{IFS} to the fieldbus and to
write received data into the \ac{IFS}. Thus, the \ac{IFS} is the
source and sink for all communication activities. Furthermore, the
\ac{IFS} acts as a temporal firewall that decouples the local
transducer application from the communication activities. A temporal
firewall~\cite{nossal:97} is a fully specified interface for the
unidirectional exchange of state information between a
sender/receiver over a time-triggered communication system. The
basic data and control transfer of a temporal firewall interface is
depicted in Figure~\ref{fig:temporalfirewall}, showing the data and
control flow between a sender and a receiver. The \ac{IFS} at the
sender forms the output firewall of the sender and the \ac{IFS} of
the receiver forms the input firewall of the receiver. Thus, small
timing errors at the sender do not propagate through the
communication channel (significant timing errors are recognized as a
failure).

\paragraph{Flow control using information push and pull paradigms.}

The sender deposits its output information into its local \ac{IFS}
according to the information {\em push} paradigm, while the receiver
must {\em pull} the input information out of its local \ac{IFS}
(non-consumable read)~\cite{elmenreich:01}. In the information push
model the sender presses information on the receiver. It is ideal
for the sender, because the sender can determine the instant for
passing outgoing information to the communication system. The
information pull model is ideal for the receiver, since tasks of the
receiver will not be interrupted by incoming messages. The transport
of the information is realized by a time-triggered communication
system that derives its control signals autonomously from the
progression of time. The instants when typed data structures are
fetched from the sender's \ac{IFS} and the instants when these typed
data structures are delivered to the receiver's \ac{IFS} are common
knowledge of the sender and the receiver. A predefined communication
schedule defines time, origin and destination of each communication
activity. Thus, the \ac{IFS} acts as a {\em temporally specified
interface} that decouples the local transducer application from the
communication task.

\begin{figure}[htb]
  \centerline{\includegraphics[width=.6\columnwidth]{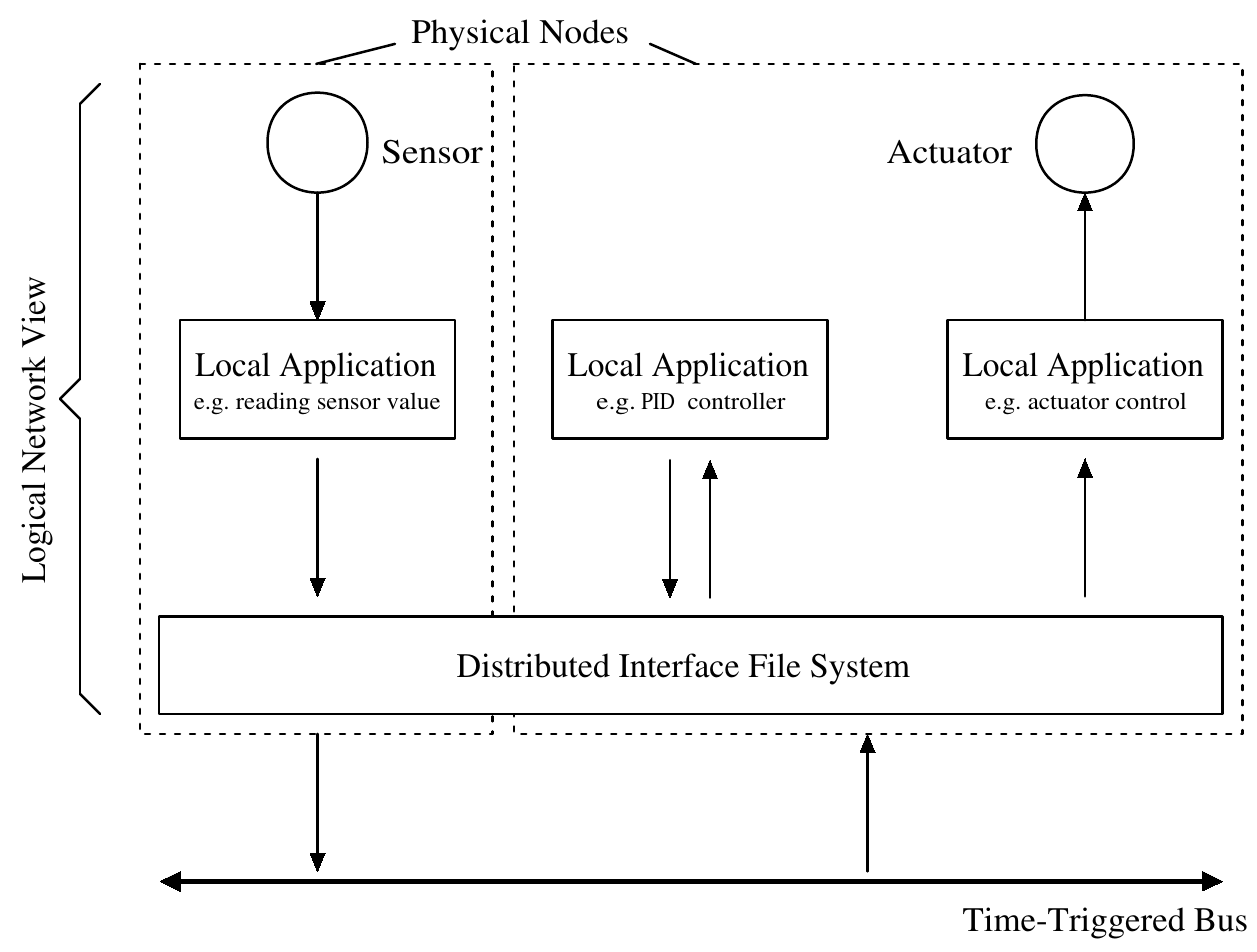}}
  \caption{Logical network view of a distributed smart transducer application}
  \label{fig:logische_netzstruktur}
\end{figure}

\subsection{Interface Separation}

\begin{figure}[htb]
  \centerline{\includegraphics[width=.5\columnwidth]{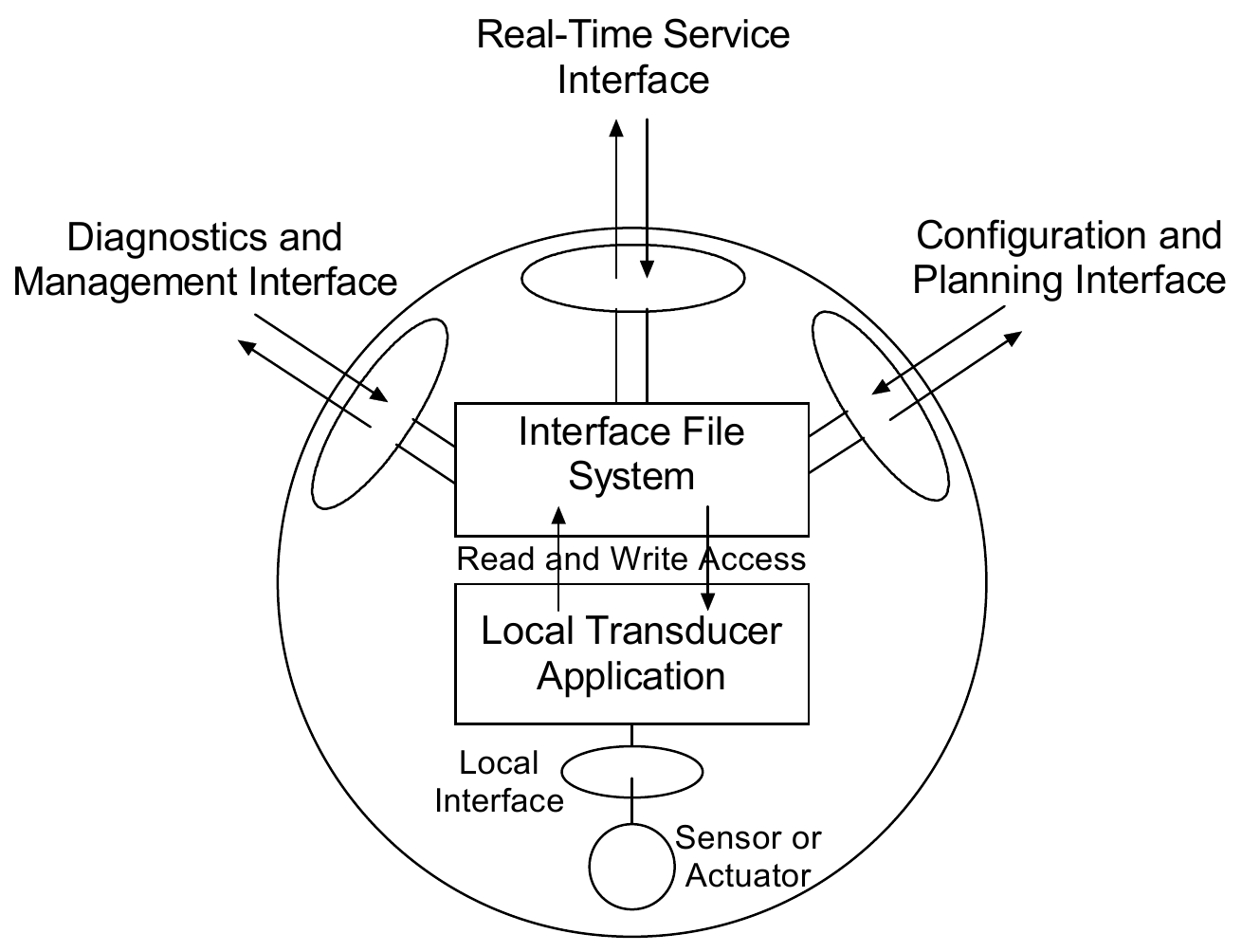}}
  \caption{The three interface types to a Smart Transducer Node}
  \label{fig:ifsnode}
\end{figure}

If different user groups access the system for different purposes,
they should only be provided with an interface to the information
relevant for their respective purpose~\cite{ran:97}. Therefore,
interfaces for different purposes may differ by the accessible
information and in the temporal behavior of the access across the
interface. As depicted in Figure~\ref{fig:ifsnode}, the STI
specifies three different interface types to a smart transducer:

\begin{description}

\item[DM interface:] This is a {\em diagnostic and management}
interface. It establishes a connection to a particular smart
transducer node and allows reading or modifying of specific \ac{IFS}
records. Most sensors need parametrization and calibration at
startup and continuously collect diagnostic information to support
maintenance activities. For example, a remote maintenance console
can request diagnostic information from a certain sensor. The DM
interface is usually not time-critical.

\item[CP interface:] The {\em configuration and planning}
interface allows the integration and setup of newly connected
nodes. It is used to generate the \dq{glue} in the network that
enables the components of the network to interact in the intended
way. Usually, the CP interface is not time-critical.

\item[RS interface:] The {\em real-time service} interface
performs a periodic communication with predictable timing behavior
among the smart transducer nodes. Communicated data is usually
data from sensors and for actuators, but may also involve
communication to and from processing nodes. This view employs
sensors for producing periodic observations of real-time entities
in the environment. For example, a temperature sensor periodically
sends the observed and locally preprocessed sensor value to the
temporal firewall of the master. Since in a time-triggered system
the time interval between sensing the environment and presenting
the sensor value at the temporal firewall~\cite{nossal:97} of the
master is known a priori, it is possible to perform a feed-forward
state estimation of the sensor value at the sensor node in such a
way, that the delivered sensor value is a good estimate of the
real-time entity's actual state at the point in time of delivery.

\end{description}

\paragraph{Naming and addressing.}

Each transducer can contain up to 64 files in its \ac{IFS}. An
\ac{IFS} file is an index sequential array of up to 256 records. A
record has a fixed length of four bytes (32 bits). An \ac{IFS}
record is the smallest addressable unit within a smart transducer
system. Every record of an \ac{IFS} file has a unique hierarchical
address (which also serves as the global name of the record)
consisting of the concatenation of the cluster name, the logical
name, the file name, and the record name.

Besides access via the master node, the local applications in the
smart transducer nodes can also execute a clusterwide application
by communicating directly with each other.

Figure~\ref{fig:logische_netzstruktur} depicts the network view for
such a clusterwide application. Note that the actual communication
between physical nodes becomes transparent for the local
applications since they exchange their data only via the \ac{IFS}.

The \ac{IFS} of each smart transducer node can be accessed via the
RS interface, the DM interface, and the CP interface for different
purposes. All three interface types are mapped onto the fieldbus
communication protocol, but with different semantics regarding
timing and data protection.

\subsection{Fieldbus Communication Protocol} \label{sec:fieldbus}

A time-triggered transport service following the specification of
the STI has been implemented in the time-triggered fieldbus
protocol TTP/A~\cite{ttpaspec}.

The bus allocation is done by a \ac{TDMA} scheme. Communication is
organized into rounds consisting of several \ac{TDMA} slots. A slot
is the unit for transmission of one byte of data. Data bytes are
transmitted in a standard UART (Universal Asynchronous Receiver
Transmitter) format. The first byte of a round is a message from the
master called fireworks byte, since this message acts as a signal to
all nodes for triggering a communication round.

The fireworks byte defines the type of the round. The protocol
supports eight different firework bytes encoded in a message of one
byte using a redundant bit code~\cite{haidinger:2000-5} supporting
error detection.

Generally, there are two types of rounds:

\begin{description}

\item[Multipartner (MP) round:] This round consists of a configuration
dependent number of slots and an assigned sender node for each
slot. The configuration of a round is defined in a data structure
called \dq{RODL} (ROund Descriptor List). The RODL defines which
node transmits in a certain slot, the operation in each individual
slot, and the receiving nodes of a slot. RODLs must be configured
in the slave nodes prior to the execution of the corresponding
multipartner round. An example for a multipartner round is
depicted in Figure~\ref{fig:ttpa_round}.

\begin{figure}[t]
  \centerline{\includegraphics[width=.6\columnwidth]{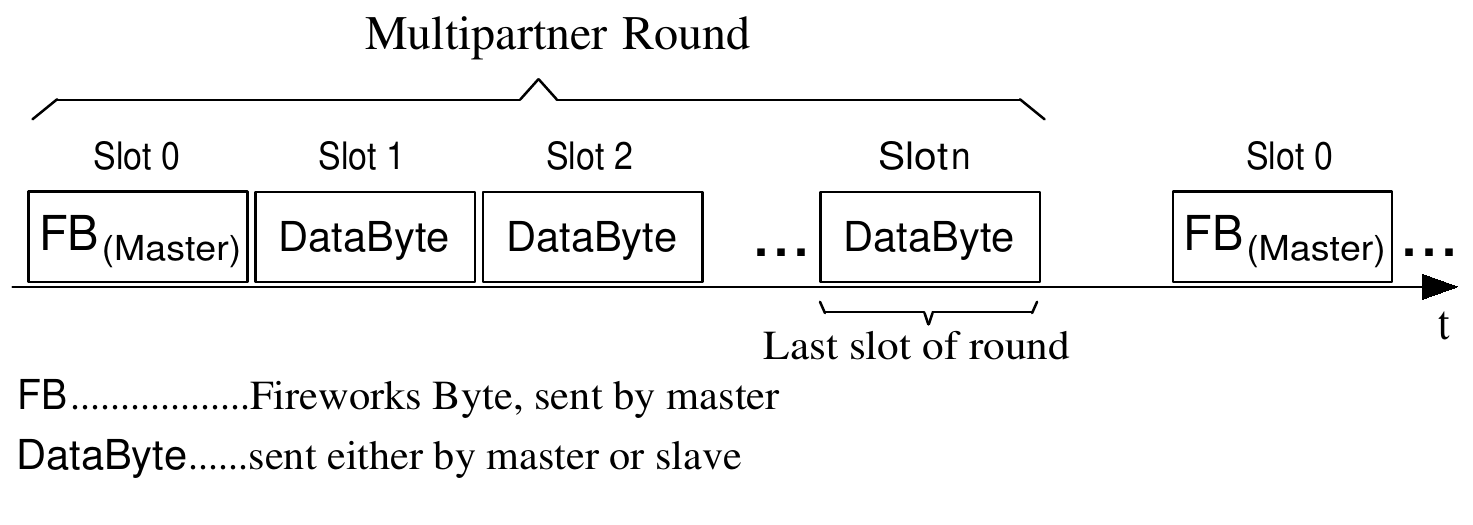}}
    \caption{A TTP/A Multipartner Round}
    \label{fig:ttpa_round}
\end{figure}

\item[Master/slave (MS) round:] A master/slave round is a special round
with a fixed layout that establishes a connection between the master
and a particular slave for accessing data of the node's \ac{IFS},
\eg the RODL information. In a master/slave round the master
addresses a data record in the hierarchical \ac{IFS} address and
specifies an action that is to be performed on that record.
Supported actions are either reading, writing, or executing a
record.

\end{description}

\begin{figure}[b]
  \centerline{\includegraphics[width=.5\columnwidth]{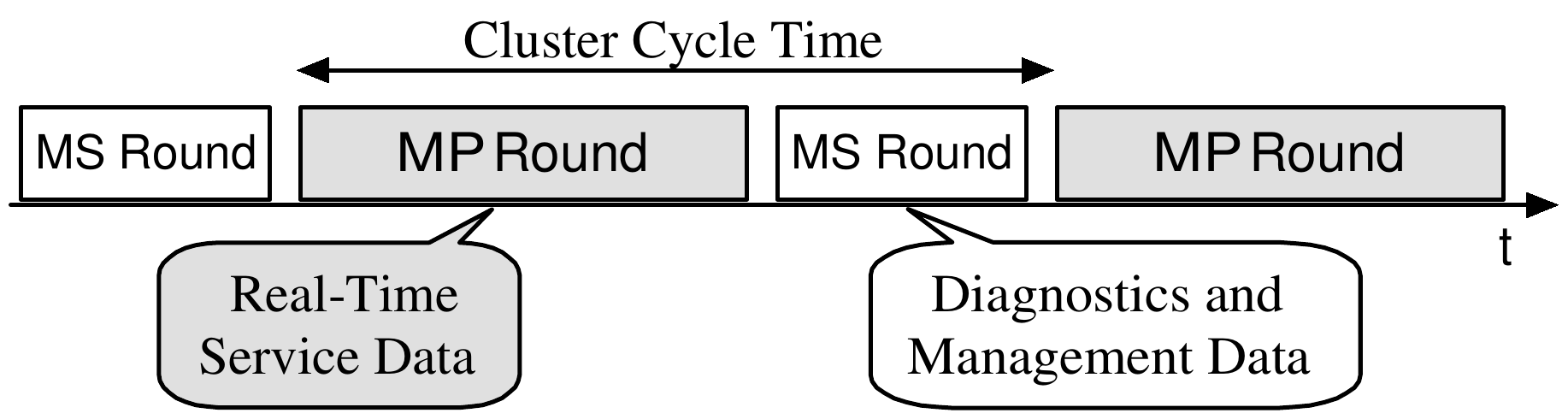}}
    \caption{Recommended TTP/A Schedule}
    \label{fig:recommended_schedule}
\end{figure}

The master/slave rounds establish the DM and the CP interface to
the transducer nodes. The RS interface is provided by periodical
multipartner rounds. Master/slave rounds are scheduled
periodically between multipartner rounds as depicted in
Figure~\ref{fig:recommended_schedule} in order to enable
maintenance and monitoring activities during system operation
without a probe effect.

\subsection{Integrating New Nodes into the Network}

New transducer nodes that are connected to a cluster must be first
configured before they can take part in the communication. A
plug-and-play configuration consists of at least three tasks: to
identify the new nodes, to obtain the documentation, and to
download the configuration.

While new node identification is trivial for many networks, it is
a difficult task in networks where deterministic behavior is
achieved by master-slave addressing. The time-triggered smart
transducer network uses a baptizing method for identification and
configuration of new nodes that does not affect the determinism of
the real-time communication of the network.

Until a logical name has been assigned to a node, it does not take
part in the multi-partner rounds. The baptize
algorithm~\cite{elmenreich:2002-5} is executed by the master to
see which nodes are connected to the {\TTPA} bus and to assign
each of them a logical name, which is unique in the current
{\TTPA} cluster.

\begin{figure}[t]
  \centerline{\includegraphics[width=\columnwidth]{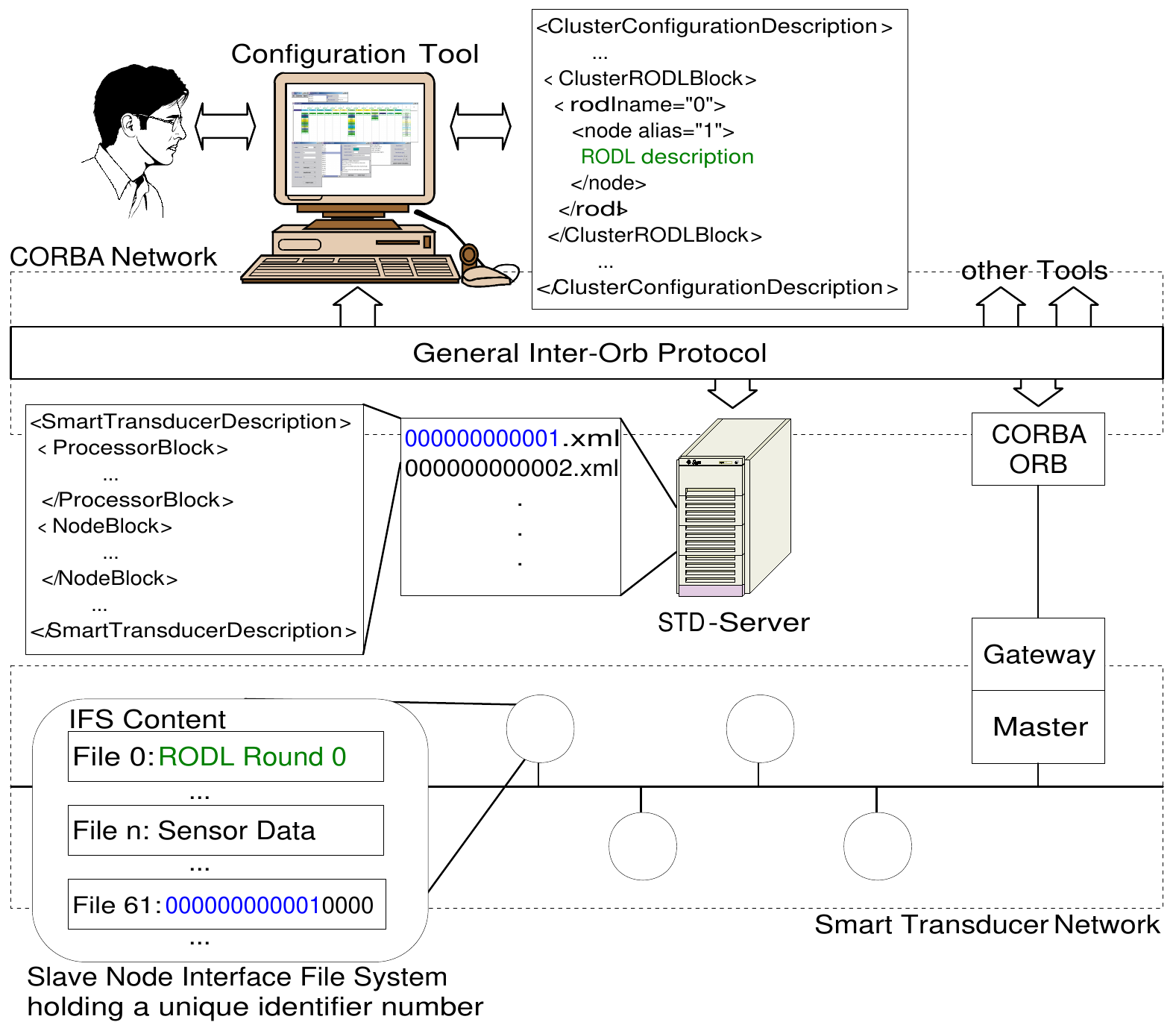}}
    \caption{Accessing a node's datasheet}
    \label{fig:datasheet}
\end{figure}

This mechanism performs a binary search on all physical node names.
A physical name is unique for every {\TTPA} node within the entire
universe of {\TTPA} nodes. The identification of a new node takes 64
iterations. After finding the unique identifier of a node, a new
logical name must be assigned to this node. The unique identifier of
a node consists of a part that describes the generic node type (the
{\em series} number) and a part that is used to distinguish between
multiple instances of a transducer type (the {\em serial} number).
The series number establishes a reference to the node's electronic
datasheet containing the necessary information for integrating the
node into the network. Datasheet information is uniformly
represented in XML (eXtended Markup Language) and can be accessed via a CORBA service. The
descriptions~\cite{pitzek:2003-6} consist of a cluster configuration
part and a smart transducer description part (cf. device description
approaches like electronic datasheets from IEEE
1452.2~\cite{ieee1451-2spec}, the IEC 62390 common automation device
profile~\cite{iec-62390}, or the Field Device Configuration Markup
Language (FDCML)~\cite{fdcml-spec}).


Figure~\ref{fig:datasheet} depicts an example for accessing a
node's datasheet via the CORBA network.

Since the configuration information is not directly stored at the
node, there is no overhead on the smart transducers themselves.

\section{Implementation Experiences}

\subsection{Smart Transducer Nodes}
\label{sec:smart_transducer_nodes}

The presented time-triggered smart transducer interface has been
implemented on several hardware platforms. The current segment of
cheap 8-bit Microcontrollers is best suited for equipping sensors
or actuators with a low-cost smart interface. We have made
experiences with node implementations on the Atmel AVR family, the
Microchip PIC and, especially for the master, the 32-bit ARM RISC
microcontrollers.

\begin{figure*}[bth]
 \centerline{\includegraphics[width=.5\textwidth]{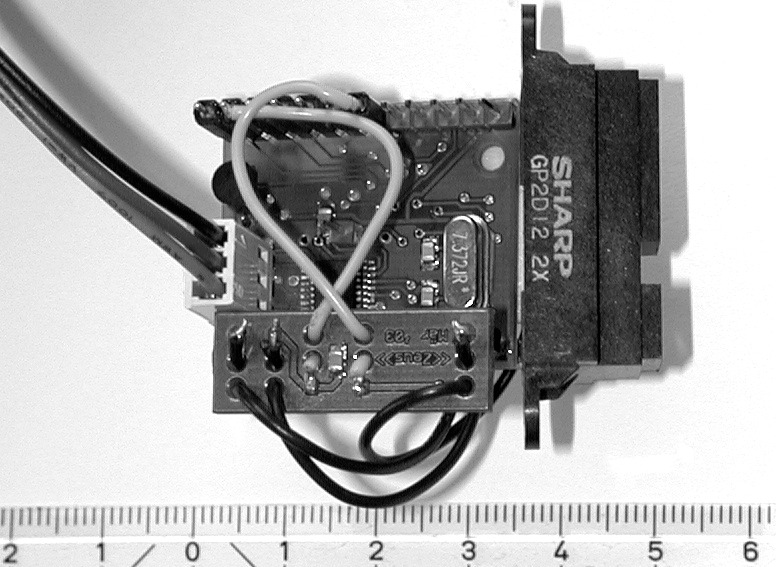}}
  \caption{Smart transducer based on Atmel 4433 microcontroller with distance sensor attached (scale in centimeter)}
  \label{fig:4433_with_distance_sensor}
\end{figure*}

Figure~\ref{fig:4433_with_distance_sensor} depicts the hardware of
a smart transducer implementation based on an Atmel AVR AT90S4433
microcontroller and an attached distance sensor. This type of
controller offers 4K Byte of Flash memory and 128 Byte of SRAM.
The physical network interface has been implemented by an ISO 9141
k-line bus, which is a single wire bus supporting a communication
speed up to 50 kBps. The wires to the left of the photo contain
the bus line and the power supply.

Table~\ref{table:ttpacomparison} gives an overview on the resource
requirements for smart transducer implementations in Atmel AVR,
Microchip PIC and ARM RISC microcontrollers.
Since the time-triggered approach follows the {\em resource
adequacy} principle~\cite[p.\,15]{kopetz:97}, the performance and
current workload at the controller does not influence the specified
real-time behavior of the network, however, a controller that
supports only a particular communication speed may not be used in
networks that specify a higher communication rate.
All three implementations held the timing requirements with a Baud
Rate of 19.2 kbps. As physical layer, an ISO 9141 k-line bus had
been used. For the Atmel AT90S4433 a maximum performance of 58.8
kbps had been tested on an RS485 physical layer.

\begin{table}[h]
  \centerline{
\begin{tabular}{|p{3.5cm}||p{2.5cm}|p{2.3cm}|p{2.3cm}|p{2.3cm}|}
 \hline
Microcontroller & Used Program Memory & Used RAM Memory & Clock Speed & Tested Baud Rate\\
 \hline
Atmel AT90S4433 & 2672B & 63B & 7.3728 MHz & 58.8 kbps\\
Microchip PIC & 2275B & 50B & 8.0 MHz & 19.2 kbps\\
ARM RISC & 8kB & n.k. & 32.0 MHz & 19.2 kbps\\
 \hline
 \end{tabular}}
 \caption{Resource requirements and
performance of time-triggered smart transducer interface
implementations
(from~\cite{troedhandl:dipl}).\label{table:ttpacomparison}}
\end{table}
The implementations on these microcontrollers show that due to the
low hardware requirements of the time-triggered smart transducer
interface it should be possible to implement the protocol on
nearly all available microcontrollers with similar features like
the Atmel or Microchip microcontroller types, that is 4KB of Flash
ROM and 128 Byte of RAM memory.

\subsection{Application Case Study}

As an example for a time-triggered smart transducer application,
an autonomous mobile robot consisting of a four-wheeled model car
with a smart transducer network for instrumenting a set of
sensors, actuators and a navigation module, has been designed and
implemented at the Vienna University of Technology.

The robot basically uses its sensors to locate objects in the
driving direction and sets its steering and speeding actuators in
order to pass the obstacles.

\begin{figure}[thb]
 \begin{center}
  \newlength\breite
  \setlength{\breite}{.7\textwidth}
  \includegraphics[width=\breite]{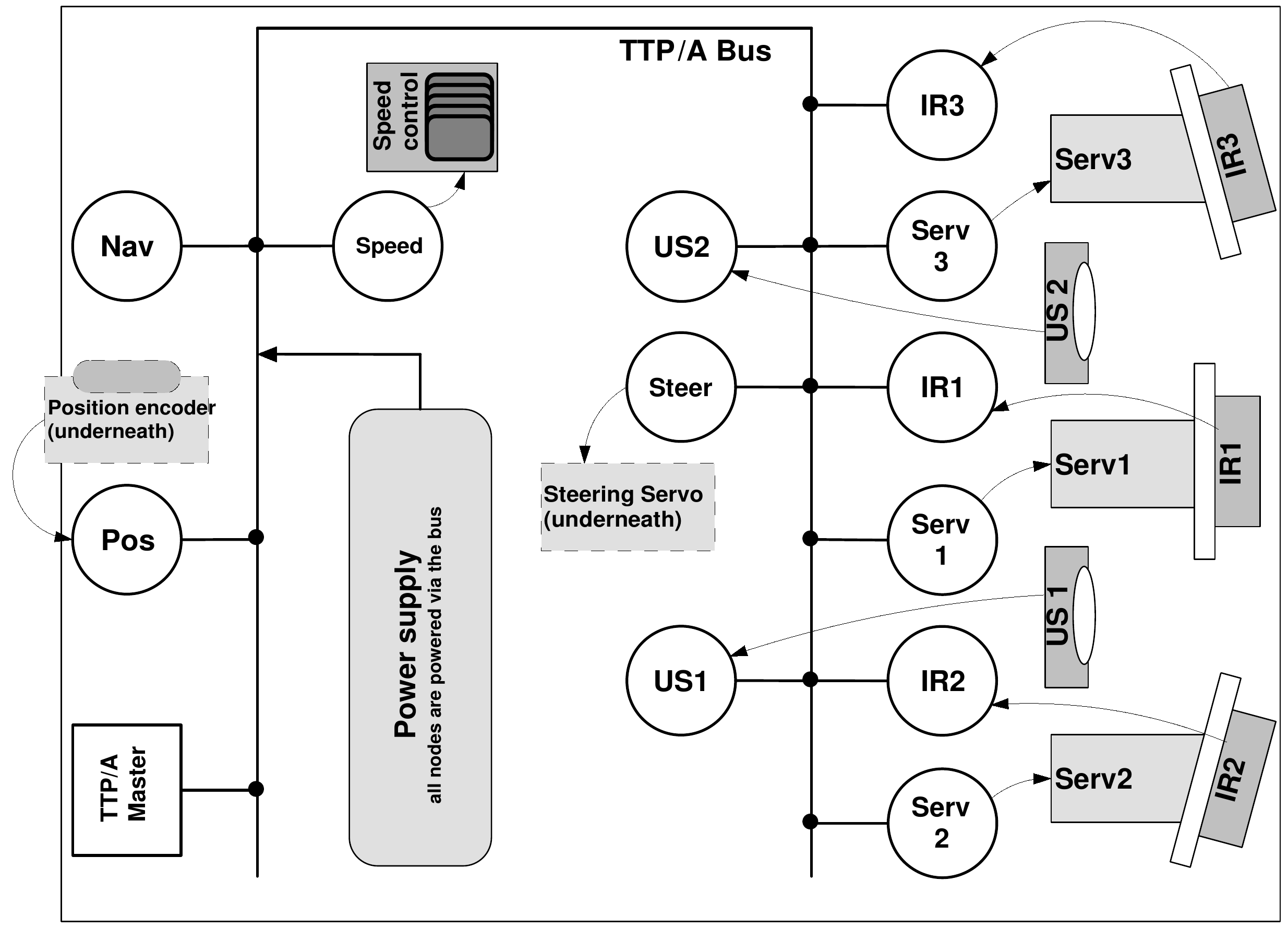}
  \footnotesize
  \begin{tabular}{p{.47\breite}p{.47\breite}}
  & \\
  IR1 \dotfill Middle infrared sensor        & US1 \dotfill Right ultrasonic sensor\\
  IR2 \dotfill Right forward infrared sensor & US2 \dotfill Left ultrasonic sensor\\
  IR3 \dotfill Left forward infrared sensor  & Pos \dotfill Position encoder sensor\\
  Serv1 \dotfill Servo for IR1               & Speed \dotfill Speed control actuator\\
  Serv2 \dotfill Servo for IR2               & Steer \dotfill Steering control actuator\\
  Serv3 \dotfill Servo for IR3               & Master \dotfill Synchronization and gateway\\
  Nav \dotfill Coordination and navigation & \\
  \end{tabular}
 \end{center}

  \caption{Smart transducer nodes on the autonomous robot}
  \label{fig:smartcar_architecture}
\end{figure}

Figure~\ref{fig:smartcar_architecture} depicts the layout of the
smart transducer nodes employed in the smart transducer network.
Each of the 6 sensor nodes and 5 actuator nodes is implemented
using an Atmel AT90S4433 microcontroller. Due to the larger memory
requirements of the master/gateway and navigation applications,
the master and navigation nodes are implemented on the more
powerful microcontrollers AT90S8515 and ATmega128 from the Atmel
AVR series.

The smart transducer network has to fulfill several real-time
tasks:

\begin{itemize}

\item The three infrared distance sensors IR1, IR2, and IR3 are
mounted on servos Serv1, Serv2, and Serv3, which swivel the
sensors for scanning the area in front of the car. Therefore,
whenever a measurement from a distance sensor is read, the
corresponding position of the respective servo must be exactly
known in order to process the measurement correctly. The
time-triggered schedule supports this coordination of servo
position and measuring time and allows to minimize the time for a
sensor sweep.

\item The two ultrasonic sensors US1 and US2 are active sensors,
\ie they emit an ultrasonic ping when performing a measurement.
Since both sensors face the same direction (the front of the car),
the measurements must be coordinated in order to avoid mutual
interference from the ultrasonic pings. This requirement can be
conveniently fulfilled by defining an appropriate phase offset in
the time-triggered schedule.

\item The control of the robot's speed and the measurement of the
covered distance are done by a feedback control loop. Using
standard control theory, the duration between measurement and
setting a new speed value has to be constant and known. The
time-triggered schedule fulfills this requirement at the outset.

\end{itemize}

%
%

Due to the predictable timing of the time-triggered system, the
communication and action schedule could be implemented very tight
and efficient, since no time is wasted for repeating messages in
case of a busy channel, etc. The whole application runs
sufficiently at a bus speed of 9600 bit/sec with a cluster cycle
of 30 ms.

Each smart transducer node was designed independently from the
overall application, most of them have been reused in other smart
transducer applications.
The robot has been used as a demonstrator for composable development
in the DSoS project (Dependable Systems of Systems, IST Research
Project IST-1999-11585). A report describing the robot
implementation in detail can be found in~\cite{dsos:pce3}.

\section{Related Work} \label{sec:relatedwork}

Real-time distributed networks for interconnection of sensors and
actuators represent a well-established research area in the
scientific community. Good overview papers on real-time
communication systems including time-triggered communication are
\cite{rushby03:buscompare,zurawski:05:ict,berge:00,jordan:95}.

However, there is not a lot of research literature that discusses
the application of hard real-time capable time-triggered network
interfaces for smart transducer networks. Most notable exceptions
are the extension of the IEEE 1451 smart transducer standard with
a time-triggered interface and, to a lesser extend, the LIN fieldbus, which has a time-triggered polling
scheme.

\subsection{IEEE 1451 with Time-Triggered Communication}

An Irish research group has developed a time-triggered smart
transducer system that incorporates the IEEE Smart Transducer
Interface Standard (IEEE 1451.2) and a Time-Triggered Controller
Area Network (TTCAN) communication protocol~\cite{doyle:02}.

The IEEE 1452 Smart Transducer Interface Standard~\cite{conway:00}
proposes a point-to-point communication interface between a Smart
Transducer Interface Module (STIM) and a Network-Capable
Application Processor (NCAP). It supports a broad range of sensor
and actuator models including those for buffered, time-triggered,
data-sequence and event-sequence models. The standard uses an
object representation for measured variables that resolves the
problem of handling physical units. Additionally, it defines the
Transducer Electronic Data Sheet (TEDS), which abstracts the
properties of sensors and actuators enabling their dynamic
definition via contextual information associated with data.

The IEEE 1451 standard comes also with a Transducer Independent
Interface (TII) that supports the multiplexing of messages, however
this TII does not support the hard real-time requirements regarding
determinism. Therefore in this project the IEEE 1451 standard has
been only partially implemented, since the TII protocol has been
replaced by the deterministic time-triggered TTCAN protocol.

\subsection{Local Interconnect Network (LIN)}

%
LIN is basically a polling protocol, where a central master issues
request messages to the slave nodes. The master node acts also as
a gateway to a higher network. The slave nodes are smart
transducers which are listening to specific messages in order to
set a control value or to send a measured value on reply. The
master issues request messages on a predefined schedule, while the
slave nodes are not aware of a global time or the current state of
the schedule. This simplifies the implementation of the slave
nodes, but does not support coordinated actions like synchronized
measurements.

Due to the polling principle (requesting a value involves the
request message, a \dq{thinking time} for the slave node and sending
the reply message), the effective bandwidth of a \ac{LIN} network
supports only applications with low bandwidth requirements, such as
less critical body electronic functions in cars.


\section{Conclusion} \label{sec:conclusion}

The static structure of time-triggered communication is an
advantage and a disadvantage at the same time. On the one hand, it
enables guaranteed deterministic timing and supports hard
real-time constraints, on the other hand, it makes it difficult to
efficiently access sparsely changing values or maintenance
facilities. In the proposed architecture for time-triggered smart
transducer networks, this problem has been overcome by a separate
implementation of virtual communication interfaces: the real-time
service interface provides the timely communication of fast
changing real-time values, like measurements or control values,
while the configuration and planning and the diagnostics and
management interfaces allows for flexible access to schedules,
sensor logs, trimming and calibration parameters, etc. Moreover,
since the time-triggered communication does not need to explicitly
address frames in their messages and avoids collisions by design,
it is much more efficient for periodic data exchange than
event-triggered or polling protocols.

The implementation of time-triggered smart transducers on several
platforms has shown that hard real-time requirements can be
fulfilled with resources of typical low-cost 8-bit
microcontrollers. Implementations of various sensors and actuators
have proven to be efficient and reusable.

Recent work in adapting standards like IEEE 1451 to time-triggered
communication interfaces has underlined the relevance and
appropriateness of the time-triggered approach for smart
transducer networks.

\setlength{\baselineskip}{14pt}

\section{Acknowledgments}

I would like to thank the anonymous reviewers for their valuable
comments on an earlier version of this paper. Furthermore thanks to
Wolfgang Haidinger and Stefan Pitzek for proofreading the paper.
This work was supported in part by the European IST project DECOS
(Dependable Embedded Components and Systems) under contract number
IST-511764 and the Austrian FIT-IT project SDSTI under contract No
808693.



\bibliographystyle{unsrt}
\bibliography{ttstn}

\begin{thebibliography}{10}

\bibitem{eccles:98}
L.~H. Eccles.
\newblock A brief description of {IEEE P1451.2}.
\newblock {\em Sensors Expo}, May 1998.

\bibitem{gait:86}
J.~Gait.
\newblock A probe effect in concurrent programs.
\newblock {\em Soft\-ware Prac\-tice and Experience}, 16(3):225--233, March
  1986.

\bibitem{mcdowell:89}
C.~E. McDowell and D.~P. Helmbold.
\newblock Debugging concurrent programs.
\newblock {\em ACM Computing Surveys}, 21(4):593--622, December 1989.

\bibitem{bengtsson:04}
M.~Bengtsson.
\newblock Condition based maintenance system technology –- {W}here is
  development heading?
\newblock In {\em Proceedings of the 17th European Maintenance Congress},
  Barcelona, Spain, May 2004.

\bibitem{par:92}
B.~Parhami.
\newblock Optimal algorithms for exact, inexact, and approval voting.
\newblock {\em Twenty-Second International Symposium on Fault-Tolerant
  Computing, 1992. FTCS-22. Digest of Papers.}, pages 404--411, July 1992.

\bibitem{elmenreich:sensorfusionintro}
W.~Elmenreich.
\newblock An introduction to sensor fusion.
\newblock Technical Report 47/2001, Technische Universit{\"a}t Wien, Institut
  f{\"u}r Technische Informatik, Vienna, Austria, 2001.

\bibitem{kopetz:2001-22}
H.~Kopetz and G.~Bauer.
\newblock The {Time-Triggered Architecture}.
\newblock {\em Proceedings of the IEEE}, 91(1):112 -- 126, January 2003.

\bibitem{kopetz:97}
H.~Kopetz.
\newblock {\em {Real-Time Systems:} Design Principles for Distributed Embedded
  Applications}.
\newblock Kluwer Academic Publishers, 1997.

\bibitem{poledna:diss}
S.~Poledna.
\newblock {\em Replica Determinism in Fault-Tolerant Real-Time Systems}.
\newblock PhD thesis, Technische Universit{\"a}t Wien, Institut f{\"u}r
  Technische Informatik, Vienna, Austria, 1994.

\bibitem{linspec}
Audi{ }AG, BMW AG, DaimlerChrysler AG, Motorola Inc.\, Volcano
  Communication~Technologies AB, Volkswagen AG, and Volvo~Car Corporation.
\newblock {LIN} specification and {LIN} press announcement.
\newblock SAE World Congress Detroit, \http{www.lin-subbus.org}, 1999.

\bibitem{suri:95}
N.~Suri, M.M. Hugue, and C.J. Walter.
\newblock {\em Advances in Ultra-Dependable Distributed Systems}.
\newblock IEEE Press, 1995.

\bibitem{kopetz:faulthypothesis}
H.~Kopetz.
\newblock On the fault hypothesis for a safety-critical real-time system.
\newblock In {\em Proceedings of the Automotiove Workshop San Diego}, CA, USA,
  January 2004.

\bibitem{omg:SmartTransducerInterfaceRFP}
Object Management Group (OMG).
\newblock {\em Smart Transducers Interface Request for Proposal}, December
  2000.
\newblock Available at \http{www.omg.org} as document orbos/2000-12-13.

\bibitem{omg:SmartTransducerInterface1.0}
Object Management Group (OMG).
\newblock {\em Smart Transducers Interface~V1.0}, January 2003.
\newblock Specification available at \http{doc.omg.org/formal/2003-01-01} as
  document ptc/2002-10-02.

\bibitem{holzmann:01}
H.~Kopetz, M.~Holzmann, and W.~Elmenreich.
\newblock A universal smart transducer interface: {TTP/A}.
\newblock {\em International Journal of Computer System Science
  {\footnotesize\sl\&} Engineering}, 16(2):71--77, March 2001.

\bibitem{nossal:97}
H.~Kopetz and R.~Nossal.
\newblock Temporal firewalls in large distributed real-time systems.
\newblock {\em Proceedings of the 6th IEEE Workshop on Future Trends of
  Distributed Computing Systems (FTDCS '97)}, pages 310--315, 1997.

\bibitem{elmenreich:01}
W.~Elmenreich, W.~Haidinger, and H.~Kopetz.
\newblock Interface design for smart transducers.
\newblock In {\em IEEE Instrumentation and Measurement Technology Conference},
  volume~3, pages 1642--1647, Budapest, Hungary, May 2001.

\bibitem{ran:97}
A.~Ran and J.~Xu.
\newblock Architecting software with interface objects.
\newblock In {\em Proceedings of the 8th Israeli Conference on Computer-Based
  Systems and Software Engineering}, pages 30--37, 1997.

\bibitem{ttpaspec}
H.~Kopetz{ }et{ }al.
\newblock Specification of the {TTP/A} protocol.
\newblock Technical report, Technische Universit{\"a}t Wien, Institut f{\"u}r
  Technische Informatik, Vienna, Austria, September 2002.
\newblock Version 2.00.

\bibitem{haidinger:2000-5}
W.~Haidinger and R.~Huber.
\newblock Generation and analysis of the codes for {TTP/A} fireworks bytes.
\newblock Research Report 5/2000, Technische Universit{\"a}t Wien, Institut
  f{\"u}r Technische Informatik, Vienna, Austria, 2000.

\bibitem{elmenreich:2002-5}
W.~Elmenreich, W.~Haidinger, P.~Peti, and L.~Schneider.
\newblock New node integration for master-slave fieldbus networks.
\newblock In {\em Proceedings of the 20th IASTED International Conference on
  Applied Informatics (AI 2002)}, pages 173--178, February 2002.

\bibitem{pitzek:2003-6}
S.~Pitzek and W.~Elmenreich.
\newblock Configuration and management of a real-time smart transducer network.
\newblock In {\em Proceedings of the 9th IEEE International Conference on
  Emerging Technologies and Factory Automation (ETFA 2003)}, Lisbon, Portugal,
  September 2003.

\bibitem{ieee1451-2spec}
Institute of Electrical and Electronics Engineers, Inc.
\newblock {\em {IEEE} Std 1451.2-1997, Standard for a Smart Transducer
  Interface for Sensors and Actuators - Transducer to Micro-processor
  Communication Protocols and Transducer Electronic Data Sheet ({TEDS})
  Formats}, 1997.

\bibitem{iec-62390}
International Electrotechnical Commission,.
\newblock {\em {IEC} TR 62390, Common automation device – Profile guideline},
  2005.
\newblock First edition 2005-01.

\bibitem{fdcml-spec}
FDCML.
\newblock {\em {Field Device Configuration Markup Language FDCML} 2.0
  Specification}, 2005.
\newblock Version 1.0, Available at \http{www.fdcml.org}.

\bibitem{troedhandl:dipl}
C.~Tr{\"o}dhandl.
\newblock Architectural requirements for {TTP/A} nodes.
\newblock Master's thesis, Technische Universit{\"a}t Wien, Institut f{\"u}r
  Technische Informatik, Vienna, Austria, 2002.

\bibitem{dsos:pce3}
W.~Elmenreich, W.~Haidinger, H.~Kopetz, T.~Losert, R.~Obermaisser,
  M.~Paulitsch, and C.~Tr{\"o}dhandl.
\newblock Initial demonstration of smart sensor case study.
\newblock {\em DSoS Project (IST-1999-11585) Deliverable PCE3}, April 2002.

\bibitem{rushby03:buscompare}
J.~Rushby.
\newblock A comparison of bus architectures for safety-critical embedded
  systems.
\newblock Technical Report NASA/CR–2003–212161, National Aeronautics and
  Space Administration, Langley Research Center, March 2003.

\bibitem{zurawski:05:ict}
R.~Zurawski, editor.
\newblock {\em The Industrial Communication Technology Handbook}.
\newblock CRC Press, Boca Raton, FL 33431, USA, 2005.

\bibitem{berge:00}
J.~Berge and S.~Mitschke.
\newblock Building better open networks using foundation fieldbus and {OPC}.
\newblock {\em Sensors Magazine}, February 2000.

\bibitem{jordan:95}
J.~R. Jordan.
\newblock {\em Serial Networked Field Instrumentation}.
\newblock John Wiley \& Sons, 1995.

\bibitem{doyle:02}
P.~Doyle, D.~Heffernan, and D.~Duma.
\newblock A time-triggered transducer network based on an enhanced {IEEE} 1451
  model.
\newblock {\em Microprocessors \& Microsystems Journal}, December 2002.

\bibitem{conway:00}
P.~Conway, D.~Heffernan, B.~O'Mara, D.~P. Burton, and T.~Miao.
\newblock {IEEE 1451.2: An} interpretation and example interpretation.
\newblock In {\em Proceedings of the Instrumentation and Measurement Technology
  Conference}, pages 535--540, Baltimore, MD, USA, May 2000.

\end{thebibliography}

\end{document}